# Matter distribution for power spectra with broken scale invariance

Luca Amendola[1,2], Stefan Gottlöber[3], Jan Peter Mücket[3], Volker Müller[3]

[1] *Osservatorio Astronomico di Roma*
*Viale del Parco Mellini, 84*
*I-00136 Rome, Italy*

[2] *NASA/Fermilab Astrophysics Center*
*PO Box 500*
*Batavia IL 60510, USA*

[3] *Astrophysical Institute Potsdam*
*An der Sternwarte 16*
*D-14482 Potsdam, Germany*

ABSTRACT

To test the primordial power spectra predicted by a double inflationary model with a break of amplitude $\Delta = 3$ at a scale of $2\pi/k \approx 10 h^{-1}$ Mpc and CDM as dominant matter content, we perform PM simulations with $128^3$ particles on a $256^3$ grid. The broken scale invariance of the power spectra explains the extra power observed in the large-scale matter distribution. COBE-normalized spectra and a linear biasing with $b \approx 2$ are shown to reproduce the reconstructed power spectra from the CfA catalog. Identifying galactic halos with overdensity of approximately two times the cell variance, we can fit the angular correlation function using both the Limber equation and creating a APM-like angular projection with the observed luminosity function. Finally, the higher order moments of the galaxy distribution are shown to fit reasonably well the observed values.

## 1   Introduction

Inflationary cosmological models predict that the observable part of the universe is quasi-flat, i.e. the total density parameter is $\Omega_{tot} \approx 1$. Combined with the density of baryons $\Omega_{bar} \approx 0.02 h^{-2}$ ($h$ denotes the Hubble constant in units of 100 km s$^{-1}$Mpc$^{-1}$) following from the theory of primordial nucleosynthesis this requires most of matter in the Universe to be nonbaryonic. The most elaborate model of structure formation assumes an universe dominated by cold dark matter (CDM) with a Harrison–Zeldovich spectrum of initial perturbations. With a biasing $b_g \approx 2$ it has successfully explained the observed hierarchy of cosmic structures on scales smaller than approximately 10 $h^{-1}$ Mpc. However, recent observations seem to indicate that the predictions of the standard CDM model on very large scales and on small scales are incompatible (Efstathiou *et al.* 1990, Maddox *et al.* 1990, Fisher *et al.* 1993, Saunders *et al.* 1991, Vogeley *et al.* 1992).

The standard model is based on two assumptions, namely, (1) that the primordial perturbation spectrum generated during inflation is of Harrison-Zeldovich type, and (2) that the dark matter, which gives the main contribution to the density of the Einstein-de Sitter universe, is cold. In order to change the theoretical predictions of this model, one can modify either of these assumptions. In the first case, improved inflationary scenarios lead to other than Harrison-Zeldovich spectra. In the second case, widely discussed candidates for closing the universe are mixed dark matter or a cosmological term. In a recent paper, a change of the first type has been investigated (Gottlöber, Mücket, Starobinsky 1994). The underlying inflationary model shows two consecutive stages of exponential expansion with a short intermediate stage of power law expansion (Gottlöber, Müller, Starobinsky 1991). The two driving mechanisms are vacuum polarisation effects and a massive scalar field. In this case the resulting power spectra with broken scale invariance (BSI) are of Harrison-Zeldovich type only in the limit of very small and very large scales. In the intermediate range they are steeper. The spectra are characterized by the ratio $\Delta$ of the power of a Harrison-Zeldovich spectrum to the power of the BSI spectrum on small scales assuming both are normalized on large scales (COBE normalization). The scale $k_{br}$ denotes the onset of the break in the perturbation spectrum at small scales, i. e. for $k > k_{br}$ the spectrum is of Harrison-Zeldovich type. The quantity $\Delta$ mainly depends on the parameters characterising the inflationary stages (the mass of the scalar particle and the coupling constant of the higher-order terms), whereas $k_{br}$ depends on the energy density of the scalar field at the onset of inflation. The best fit to observations is reached with $\Delta$ between 2 and 3 and $k_{br}^{-1}$ in the range between $1h^{-1}$ Mpc and $4h^{-1}$ Mpc (Gottlöber, Mücket, Starobinsky 1994). In the following we adopt $\Delta = 3$ and $k_{br}^{-1} = 1.5h^{-1}$ Mpc. Starting from these spectra, we have performed N-body simulations using the particle-mesh code of Kates *et al.* (1991), extended to three dimensions. In this paper we report on the linear and non-linear clustering properties that characterize the matter in these simulations. In Section 2 we begin by discussing the spatial correlation function and the power spectrum in our simulations.

One of the most convincing evidence for rejecting the standard CDM model comes probably from the angular correlation function (ACF). Even if the information on redshift is lost, the angular surveys provide such an enormous amount of data, millions of galaxy positions, that they turn out to tightly constrain theoretical models. As it is well known, the ACF reported by Maddox *et al.* (1990) is not matched by the standard CDM model; on angular scales larger than two or three degrees, the observed ACF is found indeed to be significatively larger than predicted by CDM, when CDM is normalized to small scales. One can say that since the publishing of the APM results the minimal requirement for any new model of galaxy formation is that the correct ACF be reproduced. In Section 3 we report the test of our model against the observed ACF.

As we already stated, to match the correlation functions is only a minimal requirement, though a very significant one, for a model to be acceptable. In recent years many higher order statistical measures of the observed distribution of galaxies have been discussed and calculated. A direct extension of the correlation function are the three- and four-point correlation functions, thoroughly discussed in literature. Since their estimate is very noisy and time-consuming, often some integral version of the $n$-point correlation functions is computed from the data. Particularly simple and interesting are, for instance, the skewness and the kurtosis of the count-in-cells (e.g. Coles & Frenk 1991, Saunders *et al.* 1991, Bouchet *et al.* 1993, Gaztañaga 1992, 1994). In Section 4 we derive the higher order moments of our simulations, and compare them with the data. Finally, in Section 5 we draw our conclusions.

## 2  Power spectrum and correlation function

We have performed $N$-body simulations with $128^3$ particles in cubes with a $256^3$ grid. The simulations were made with four different box sizes, $L = 25h^{-1}$ Mpc, $75h^{-1}$ Mpc, $200h^{-1}$ Mpc and $500h^{-1}$ Mpc in order to get a resolution high enough to identify the places where galactic halos form and, for the same spectrum and with the same method, to model the large scale matter distribution (cp. Kates *et al.* 1994).

We consider a CDM model ($\Omega = 1, H = 50$km/s/Mpc) with the transfer function of Bond and Efstathiou (1984) and the primordial perturbation spectra with broken scale invariance calculated by Gottlöber, Müller, and Starobinsky (1991). We have normalized the spectra with the one year COBE result $\sigma_T(10°) = (30 \pm 7.5)\mu K/2.735K$ (Smoot *et al.* 1992). The inflationary model we are considering produces a negligible contribution of gravitational waves to the microwave anisotropy. The rms multipole values $C_l$ for our perturbation spectrum are substantially smaller than for the standard CDM model for $l > 30$ (Gottlöber and Mücket, 1993). We did not include the new analysis presented in Wright et al. (1994) which leads to a almost 10 % increase (20% following Górski *et al.* 1994) of the large-scale normalization. Taking this into account, the biasing factor calculated from the spectra would decrease by the same amount.

The linear analysis of the power spectrum (Gottlöber, Mücket, Starobinsky 1994) has shown that we need a bias $b = 2.18$ for transforming structures of dark matter particles into luminous matter. The power spectrum one actually computes from a finite size, finite resolution simulation is an acceptable approximation of the theoretical one only in a range of wavelengths $k$ for which $2\pi/L \ll k \ll 2\pi/l$, where $l$ is the cell size and $L$ is the box size. Then, to extend the range of validity of the power spectrum, we estimate the true $P(k)$ by joining four power spectra for different box sizes (with $L = 256l$). We take the highest $P(k)$ for any value of $k$, since the effect of both finite size and finite resolution is generally to reduce the power amplitude. The reconstructed power spectrum coincides with the linear one at very large scales. At scales smaller than $k_{nl} \approx 0.2h$ Mpc$^{-1}$ the BSI model shows more power than the linear spectrum and has a slope of $k^n$ with $n \approx -1.3$. Consequently, also the variance at 8 $h^{-1}$ Mpc increases. Therefore, the biasing factor defined as $b = \sigma_8^{-1}$ decreases. From the simulations we found an optimal value of $b \approx 1.7$ (instead of the linear 2.18).

To compare the reconstructed spectrum with the spectrum of the CfA catalogue we transform it into redshift space using the Kaiser (1984) and Peacock (1991) corrections. The resulting spectrum is shown in Fig. 1, where we have assumed a biasing factor $b = 1.7$. For comparison the spectrum of a simulation with the standard CDM spectrum is also included in this figure ($b = 0.9$). Note that in our simulation the highest scale mode ($k_{min}$) is overestimated by about 20 % due to the chosen representation of the spectrum in $k$-space. The maximum value of the calculated spectrum would really be at $k \approx 0.04h$ Mpc$^{-1}$, and the spectrum will be bent down at small $k$'s as it is indicated by the data (though with very large error bars).

We identify galaxies in our simulations by means of the peak-background split formalism. Let $\sigma^2$ be the density fluctuation variance on the grid cell of a given simulation, and $\nu\sigma$ the selection threshold: only particles residing in regions with density contrast $\delta > \nu\sigma$ are identified with galaxies. Also, let $\xi$ be the correlation function of all particles, and $\xi_\nu$ the correlation function relative to the galaxies. The relation between the linear biasing $b \equiv (\xi_\nu/\xi)^{1/2}$ and the threshold $\nu$ is quite complicated, because it involves the full probability distribution of the underlying fields, which is in general unknown. The simple formula given in the classical paper by Kaiser (1984) cannot be applied here because it holds only if one identifies each region above threshold with a single galaxy, while we put the density field in the regions above threshold equal to the underlying density field. Moreover, the Kaiser formula holds only for Gaussian field, while the matter clustering is certainly non-Gaussian by the present. A biasing scheme closer to ours has been investigated by Catelan *et al.* (1994), who give the relation $b = b(\nu, \sigma)$ for a lognormal field, which is known to approximate the real density probability distribution. The general trend, at least in the limit of small correlation, is that $b$ increases with increasing $\nu$ and decreasing $\sigma$ (i.e., increasing the box size). As a consequence, in a larger box one has to use a smaller threshold $\nu$ to get the same amplification $b$. The lognormal relation better approximates our results: to have $b \approx 1.7$ in a field with $\sigma \approx 4.5$ (relative to the 200 $h^{-1}$ Mpc box) one needs a threshold $\nu \approx 2$. As we will show, with this threshold our model reproduces several observational features of the galaxy clustering.

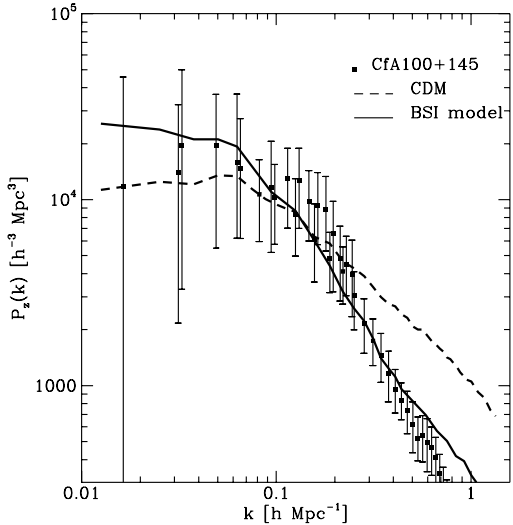

Figure 1: Comparison of simulated power spectra using initial conditions with broken scale invariance and a bias $b = 1.7$ (solid line), standard CDM model with $b = 0.9$ (dashed line) and data from the CfA catalog (Vogeley *et al.* 1992).

Finally, we have computed the spatial correlation function from the simulations. Using all particles the BSI model yields a slope of 1.6 - 1.7 with a correlation radius of $2h^{-1}$ Mpc. A biasing procedure as described above changes the slope to 1.7 - 1.8, and the correlation radius increases to $(4.5 - 5.5)h^{-1}$ Mpc (depending on the box length). In our simulation the standard CDM model yields a slope of 2.0 which is too steep, while its correlation radius is $5h^{-1}$ Mpc for all particles. In Fig. 2 we present the spatial correlation functions for our BSI model.

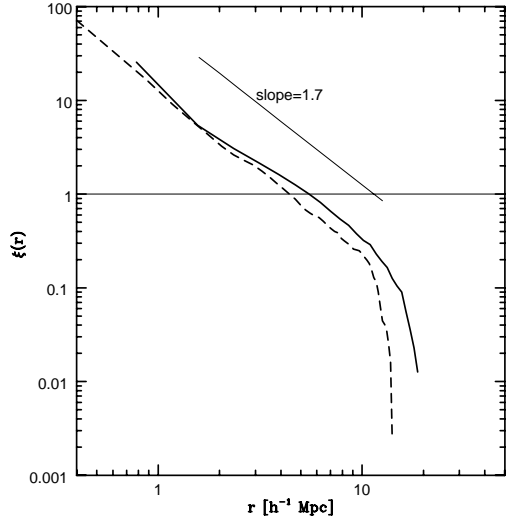

Figure 2: Two-point correlation function in simulations of BSI spectra in the 75 $h^{-1}$ Mpc (dashed line) and $200h^{-1}$ Mpc (solid line) box for particles in cells with overdensity $\nu = 2.1$ and $\nu = 1.8$, respectively.

## 3    The angular correlation function

We estimate the angular correlation function in two ways: indirectly, from the power spectrum of the $N$-body simulation, and directly, performing a projection of the $N$-body particles on a portion of sphere according to the luminosity function (Coles *et al.* 1993, Moscardini *et al.* 1993).

From the knowledge of $P(k)$, one can derive the ACF $w(\theta)$ via the Limber equation:

$$w(\theta) = \frac{\int_0^\infty kP(k)dk \int_0^\infty y^4\phi^2(y)J_0(k\theta y)dy}{2\pi^2(\int_0^\infty y^2\phi(y)dy)^2}, \qquad (1)$$

where $\phi(y) = y^{-0.5}\exp\left[-(y/y^*)^2\right]$ is the selection function. Inserting the linear power spectra into Eq. (1) and scaling the angular correlation function to the Lick depth ($y^* \approx 240h^{-1}$ Mpc) one can compare the theoretical predictions with the observational data for $w(\theta)$ from the APM survey (Maddox *et al.* 1990). However, the use of linear approximation in the calculation of $w(\theta)$ is justified for large $\theta$ only. For angles up to $\theta \sim 10°$, the agreement of the theoretical predictions with the data is very good, if spectra with $\Delta = 3$ and $k_{br}^{-1} = (1...4)h^{-1}$ Mpc are considered (Gottlöber, Mücket, Starobinsky 1994). The nonlinear behaviour can be described by using the power spectra reconstructed from the different boxes. As already mentioned on nonlinear scales the slope becomes $P(k) \propto k^{-1.3}$ so that the theoretical angular correlation $w(\theta)$ matches the observed small-angular behaviour $w(\theta) \propto \theta^{-0.7}$ very well as long as the finite cell size of the smallest simulation box does not smear out the correlations. The unbiased angular correlation function calculated in this way are shown in Fig. 4a and 4b as dotted lines. In the case of the BSI model the line can be shifted to match the data assuming a biasing as introduced above.

We have also computed the ACF by projecting the particles onto a spherical surface 90° wide and estimating $w(\theta)$ directly from the angular map. The details of this procedure have been already given in literature (e.g. Coles *et al.* 1993, Moscardini *et al.* 1993), so we only sketch the method here. We have calculated the ACF both for all particles and for particles selected by a density threshold as reported in Section 2. We have replicated the simulation box by reflection in such a way that it covers a cone with opening 90° and depth $\sim 600 h^{-1}$ Mpc, so as to reproduce the observational cone of the Lick catalog, to which the APM data have been scaled. We need three and eight levels for boxes of $L = 200 h^{-1}$ Mpc and $L = 75 h^{-1}$ Mpc, respectively, to contain the Lick cone (we did not consider the smallest and the biggest simulations here). We need the replication procedure to get both the necessary resolution on small scales and the Lick depth. However, by this method we get also a small systematic at $\theta > 10°$, as can be seen by testing the procedure with a Poisson distribution. Therefore, we have restricted our calculations to $\theta < 8°$. On small scales we are limited by the resolution of the simulations so that we considered only $\theta > 0.3°$.

According to the luminosity function we assign an absolute magnitude to each particle through a random process. From the distance to the observer, located in the center of the 'original' box, we derive the apparent magnitude $m$. If $m \leq m_{Lick} = 18.4$ (see Maddox *et al.* 1990), the particle is projected on the surface. In Fig. 3 we show the corresponding distribution of particles assuming a density threshold of 3 particles per cell (see below). Since the largest contribution comes from the box in which the observer is situated, and from the box directly on top of it, we expect that the effect introduced by the box replication is a minor one, at least for not very large separations. To test this, we have also calculated the ACF projecting only particles with brighter limiting apparent magnitudes, $m_0 = 16.4$ and $m_0 = 14.4$. In this way the characteristic depth $D^* = \text{dex}[-0.2(m_0 - M^* - 25)]$ becomes much smaller, and the effect of box duplication is greatly reduced. Then, we shift the ACF to the Lick scale by means of the scale dependence contained in the Limber equation (see e.g. Groth & Peebles 1977, Peebles 1980, Maddox *et al.* 1990). Our tests have confirmed the reliability of the method of replication.

The estimator we use for the ACF is

$$w(\theta) = F \frac{CC}{CR} - 1, \qquad (2)$$

where $CC$ is the number of pairs in the real angular catalog at angular separation $\theta$, $CR$ is the number of pairs in the crossed real-random catalogs, and $F$ is the ratio of densities of the random and the real catalog ($F > 1$ to reduce the Poissonian noise). The results for the discussed models are shown in Fig. 4a (standard CDM model) and 4b (BSI model). The squares denote the angular correlation function for all particles. While the standard CDM simulation clearly cannot fit the angular correlation function, we find a good agreement with the dashed curves calculated from the reconstructed power spectrum. The scattering of the data is due to measurements in different directions and measurements in different simulation boxes.

To introduce biasing in our BSI simulations, we follow the procedure described in Section 2. For retaining the full spatial resolution we identify galactic halo candidates on the grid by imposing discrete thresholds without any smoothing. Assuming a threshold of 3 particles per cell (200 $h^{-1}$ Mpc, $\sigma$=4.5) or 4 particles per cell (75 $h^{-1}$ Mpc, $\sigma = 7.5$), and projecting the galaxies according to the galaxy luminosity function, we get reasonable angular correlation functions in the range of $0.3^0 < \theta < 8^0$ (triangles in Fig. 4b). In comparison with the CDM model, the selected galaxies show clearly the required extra power at separations larger than $2^0$. The thresholds correspond to values of $\nu \approx 2.1$ and 1.8 for the smaller and larger box sizes, respectively.

Figure 3: Angular projection of particles in cells with overdensity $\nu = 2.1$. This map should be compared with the APM data.

## 4   Higher-order moments

If the distribution of galaxies is not Gaussian, the two-point correlation function does not fully characterize the clustering properties. Even if the initial distribution was Gaussian, as our inflationary model predicts, non-Gaussianity is induced by the non-linear gravitational effects. Indeed, the deviations from Gaussianity can be expanded in a perturbative series of the variance of the fluctuation field (Juszkiewicz $et\ al.$ 1993, Kofman & Bernardeau 1994). Theory and observations agree in finding traces of non-Gaussianity up to very large scales, where $\xi(r) \ll 1$.

A particularly simple measure of non-Gaussianity is provided by the higher order moments of the counts in cells. Let us denote with $n_i$ the number of galaxies in the $i$-th cell in a partition of a given volume in $N$ cells, and with $\hat{n}$ its mean. Let $\mu_m$ be the dimensionless central moment of order $m$, $\mu_m = <(n_i - \hat{n})^m>/\hat{n}^m$, and let $\kappa_m$ be the corresponding dimensionless cumulant (or connected moment). To the first few orders the relation between $\mu_m$ and $\kappa_m$ is: $\kappa_2 = \mu_2$, $\kappa_3 = \mu_3$, $\kappa_4 = \mu_4 - 3\mu_2^2$, etc. (see, e.g., Cramer 1966). For a Gaussian field, $\kappa_m = 0$ for $m > 2$. Since the galaxies are supposed to be a discrete sampling of an underlying continuous field, we should subtract from the cumulants the so-called shot-noise terms, i.e. the corresponding moments of a Poissonian distribution. This is actually just a hypothesis, because the galaxies could be far from being a Poissonian sampling of the field, but this is what has been routinely adopted in the data elaboration. In any case, it should be important only in the limit of very large scales, when the distribution will tend to be effectively Poissonian, and of very small scales, when the discrete nature of galaxies dominate over the clustering. Let us notice another point about the moment estimators $\kappa_m$. It is well known that when the moments of a distribution are evaluated from a sample, they are biased estimators of the true moments (see e.g. Cramer 1966). They are good estimators only in the limit of $N \to \infty$, if $N$ is the number of independent measures (i.e. the size of the sample). We will oversample the simulation box with a large number of randomly placed cells (of order $10^4$ for the smallest cells, down to $10^3$ for the largest ones), so that the $1/N$ correction should not be relevant: however, our cells are clearly not independent, so the biasing of the estimators can slightly affect our results (as well as the data) on very large scales.

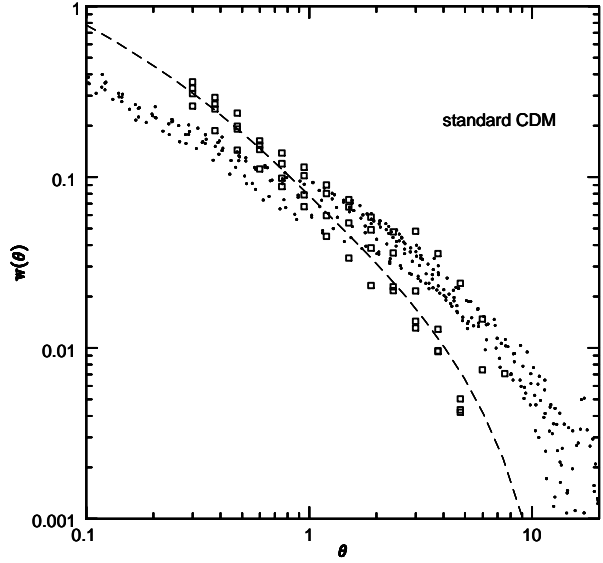

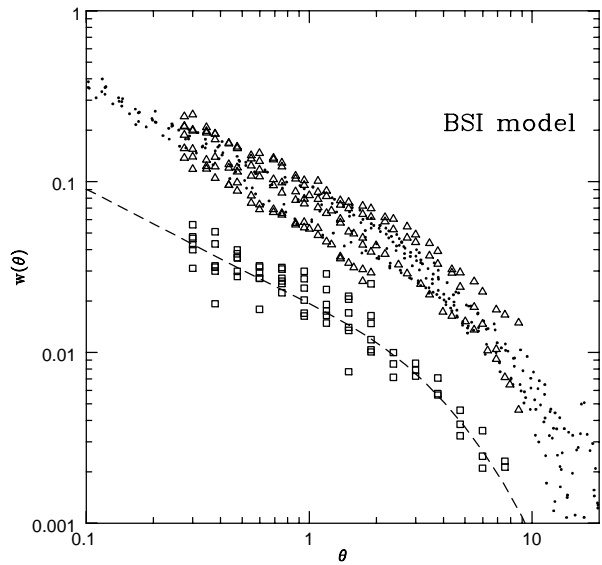

Figure 4: Comparison of the APM angular correlation function (Maddox *et al.* 1990, dots) with the simulations: *a)* CDM spectrum (squares); *b)* spectra with broken scale invariance, both for all particles (squares) and for the biased particles (triangles). The dashed lines are the correlation functions calculated from the Limber equation.

Generally speaking, the cumulant $\kappa_n$ of a fluctuation field filtered through a window $W(R)$ with characteristic scale $R$ is related to the $n$-point reduced correlation function $\xi_n$ via the integral equation

$$\kappa_m(R) = V_R^{-m} \int_{V_R} \left[ \prod_{i=1}^m d^3 x_i W_R(x_i) \right] \xi_n(x_1, ... x_m). \tag{3}$$

Through Eq. (3), a model for the higher-order correlation functions is reflected in a model for the cumulants $\kappa_n$. The most popular of such models is probably the hierarchical scaling, according to which any function $\xi_n$ is proportional to products of $n-1$ two-point functions summed over all possible distinct tree graphs. Then, the following simple scaling relation is established among the cumulants $\kappa_n$:

$$\kappa_m = S_m \kappa_2^{\eta_m}, \tag{4}$$

where $\eta_m = m - 1$ and where the $S_m$ are constants (except for a weak scale dependence introduced by the data windowing). On scales much larger than the correlation length of the fluctuation field, the scaling relation holds for general random fields at the lowest non-trivial order in the variance. It is interesting that it seems to give a fairly good description of data down to quite small scales, perhaps even in the strongly non-linear regime (although possibly with a different set of $S_m$'s). On the other hand, higher-order perturbation theory in gravitational clustering (along with assumption of initial Gaussianity) leads to a prediction of the constants $S_m$ in the small variance regime, i.e. at large scales (Peebles 1980, Juszkiewicz & Bouchet 1992, Bernardeau 1992, 1994). After smoothing, the constants $S_m$ can be estimated as function of the linear power spectrum. Their value can depend also on the bias mechanism (Fry & Gaztañaga 1993), and marginally on the redshift space distortion (Fry & Gaztañaga 1994).

Observationally, the scaling relation describes the data of several surveys from a few megaparsecs to $50h^{-1}$ Mpc and beyond (Saunders *et al.* 1991; Loveday *et al.* 1992; Bouchet *et al.* 1993; Gaztañaga 1992, 1994). Values for $S_m$ up to $m = 9$ are available in literature. At large scales they are consistent with the gravitational perturbation theory; the errorbars however are quite large, especially for $m > 4$. In $N$-body simulations, the agreement with the gravitational perturbation theory is very good on large scales, while on small scales the effects of finite volume and of discreteness make the results more controversial (e.g. Lahav 1993, Colombi *et al.* 1993).

We determined the second, third and fourth order cumulant for our models, in real space. We compared the variance/scale relation, the skewness/variance relation and the kurtosis/variance relation with the above mentioned observational data, in particular with the Stromlo-APM data of Loveday *et al.* (1992) and the APM analysis of Gaztañaga 1994 (G94). The results are plotted in Fig. 5. The errors represent the scatter between eight sets of $10^3 - 10^4$ randomly placed cells. The open circles refer to the simulation box of $75\ h^{-1}$ Mpc (selection at $\nu = 2.1$), the filled circles to the $200\ h^{-1}$ Mpc box ($\nu = 1.8$). The results merge one onto the other, from 2 to $50\ h^{-1}$ Mpc, and correctly reproduce the observations. However, the hierarchical coefficients $S_3, S_4$ for the small box decline below the ones from the large box on the larger scales. This is probably a manifestation of the boundary effects: the moments estimated from a simulation with a fixed number of particles are systematically smaller than the moments of the parent distribution (Colombi *et al.* 1993). It is interesting to observe that the scaling constants seem to be arranged in a plateau on small (non-linear) scales, while they show a continuous decrease on large scales. The decrease is due to the fact that the effective slope of the power spectrum increases for large scales. Indeed, in Fig. 6 we plotted the expected $S_3, S_4$ for top-hat cells from gravitational perturbation theory (Bernardeau 1994, scaling the spherical cells to our cubic cells) and compared with the results from the matter distribution (i.e., without bias) in the $200\ h^{-1}$ Mpc box, on scales from 5 to $50\ h^{-1}$ Mpc. The agreement with the simulation results is reasonable in the small-variance regime; the residual discrepancy is maybe due to the conversion to cubic cells, and to finite-volume effects.

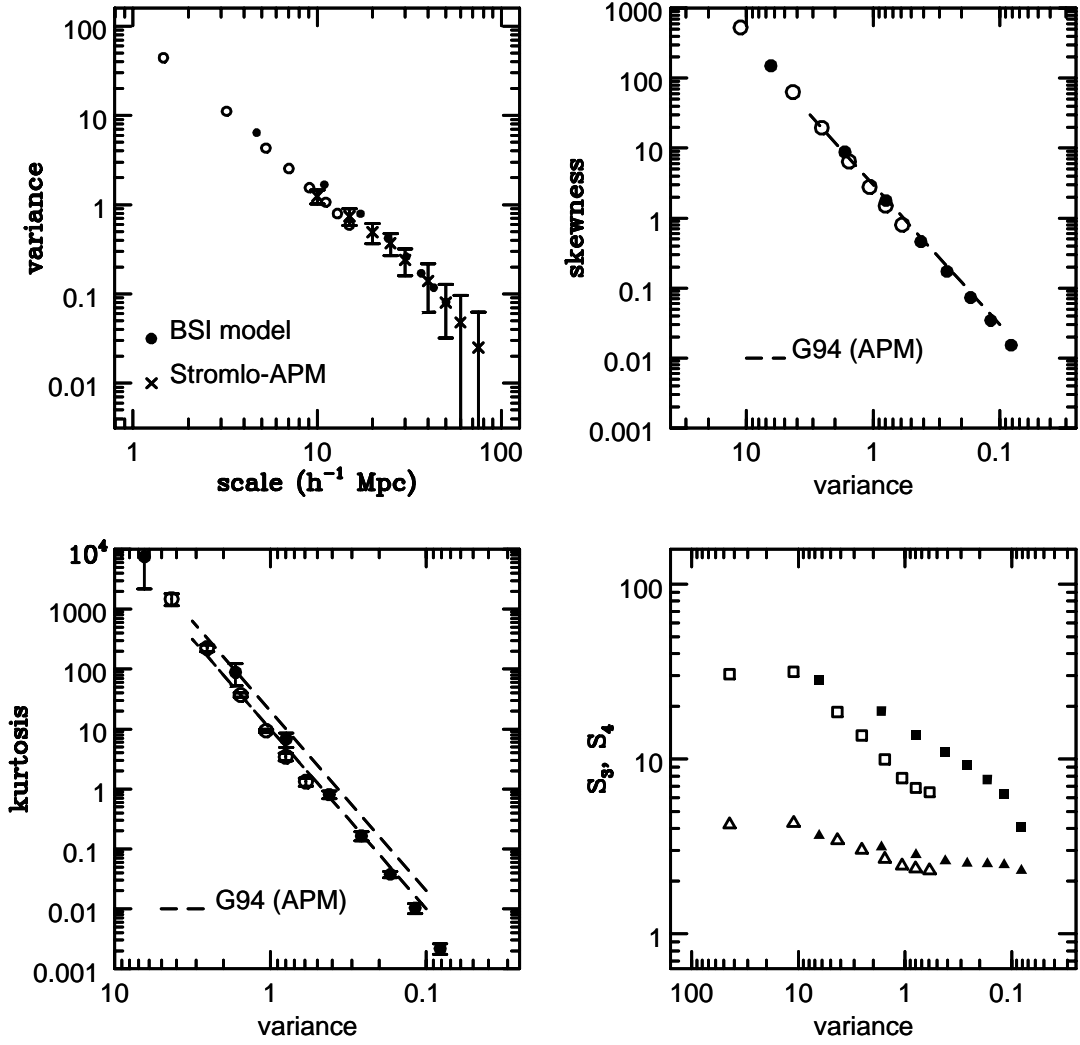

Figure 5: Higher order moments for the BSI simulation with threshold $\nu \approx 2$ in real space. Open circles represent data from the 75 $h^{-1}$ Mpc box, filled circles from the 200 $h^{-1}$ Mpc box. Clockwise from top left: variance *vs.* scale, compared with the Stromlo-APM data of Loveday *et al.* (1992) (in this plot and in the following one the errors on our data are smaller than the symbol size); skewness *vs.* variance, compared with the APM data of Gaztañaga (1994, G94; dashed line); kurtosis *vs.* variance, compared with the results from G94 which encompass scales from a few Mpc to $20h^{-1}$ Mpc; and scaling coefficients $S_3$ (triangles) and $S_4$ (squares) *vs.* variance (open symbols for the small box, filled symbols for the large box).

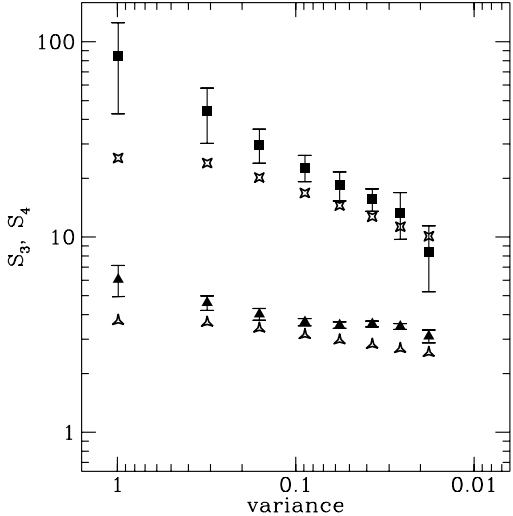

Figure 6: Scaling coefficients $S_3$ (triangles) and $S_4$ (squares) *vs.* variance relative to *all* particles in a $200h^{-1}$ Mpc box, compared to the $S_3, S_4$ expected from gravitational perturbation theory (starred symbols).

## 5  Conclusions

This paper reports on the linear and non-linear results of an extensive set of $N$-body simulations with broken scale invariance power spectrum. To our knowledge, this is the first $N$-body simulation with a primordial power spectrum arising from a double inflationary model. The results make us believe that this broken scale invariance model is an interesting addition to the list of models that help in reconciling CDM with observations. We compared our simulations with various observations, such as spatial and angular correlation functions, power spectra, higher-order moments, and we found reasonable agreement. To extend the dynamical range, we joined the results from simulation boxes ranging from $25h^{-1}$ Mpc to $500h^{-1}$ Mpc. This allowed us to fit, for instance, the observed power spectrum over almost two decades in wavelenght, and we could determine the scaling coefficients $S_3, S_4$ over three decades in the variance. The replication method we used to obtain a APM-like angular projection has been successfully tested via the limiting magnitude shifting (although we found the method unsatisfactory for $\theta \geq 10^0$).

The BSI power spectrum we investigated contains, in addition to the overall normalization, two additional parameters, the height and the location of the break. In the double inflation model, the break height is connected with the ratio of the two mass parameters characterizing the two inflationary stages. The break scale is sensibly dependent on the initial value of the scalar field. In Gottlöber, Müller, Starobinsky (1991) it was argued that this initial value may arise naturally as a quantum fluctuation. Clearly this requires further elaboration. Here our aim was to show that the values of the two parameters which were derived from linear analysis lead to non-linear clustering properties in accordance with observations.

### ACKNOWLEDGMENTS

We are grateful to Michael Vogeley for providing us with the data sets of the CfA power spectrum and to Steve Maddox for the APM angular correlation function. Ron Kates and Jörg Retzlaff helped us in preparing the $N$-body simulations. L.A. wishes to thank the Astrophysical Institute of Potsdam for the friendly hospitality during the early stages of preparation of this work, and Stéphane Colombi for very fruitful discussions. The work of L.A. at Fermilab has been supported by DOE and NASA under grant NAGW-2381. L.A. also acknowledges CNR (Italy) for financial support. S.G. acknowledges support from Fermilab for his stay at Batavia, where the final version of this paper was prepared.